\documentclass[aps,prl,10pt,amsmath,amssymb,twocolumn,longbibliography,superscriptaddress]{revtex4-2}
\usepackage{hyperref}
\usepackage{graphicx,tikz}
\usepackage{mathtools}
\usepackage{physics}
\usepackage{array}
\usepackage{multirow}
\usepackage{orcidlink}

\usepackage{soul}  
\usepackage{float}

\def\be#1\ee{\begin{align}#1\end{align}}

\begin{document}

\title{Quantization of spin circular photogalvanic effect in altermagnetic Weyl semimetals}

\author{Hiroki Yoshida\,\orcidlink{0000-0002-4505-047X}}
\affiliation{Department of Physics, Institute of Science Tokyo, 2-12-1 Ookayama,
Meguro-ku, Tokyo 152-8551, Japan}
\author{Jan Priessnitz\,\orcidlink{0009-0004-0173-9189}}
\affiliation{Max Planck Institute for the Physics of Complex Systems, N\"{o}thnitzer Str. 38, 01187 Dresden, Germany}
\author{Libor \v{S}mejkal\,\orcidlink{0000-0003-1193-1372}}
\affiliation{Max Planck Institute for the Physics of Complex Systems, N\"{o}thnitzer Str. 38, 01187 Dresden, Germany}
\affiliation{Max Planck Institute for Chemical Physics of Solids, N\"{o}thnitzer Str. 40, 01187 Dresden, Germany}
\affiliation{Institute of Physics, Czech Academy of Sciences, Cukrovarnick\'{a} 10, 162 00 Praha 6, Czech Republic}
\author{Shuichi Murakami\,\orcidlink{0000-0002-2033-9402}}
\affiliation{Department of Physics, Institute of Science Tokyo, 2-12-1 Ookayama,
Meguro-ku, Tokyo 152-8551, Japan}
\affiliation{Department of Applied Physics, the University of Tokyo, 7-3-1 Hongo, Bunkyo-ku, Tokyo 113-8656, Japan}
\affiliation{International Institute for Sustainability with Knotted Chiral Meta
Matter (WPI-SKCM$^{\mathit2}$), Hiroshima University, Higashi-hiroshima, Hiroshima
739-0046, Japan}
\affiliation{Center for Emergent Matter Science, RIKEN, 2-1 Hirosawa, Wako, Saitama 351-0198, Japan}

\date{\today}

\begin{abstract}
    We theoretically predict a spin-current analog of the quantized circular photogalvanic effect in Weyl semimetals. This phenomenon is forbidden in antiferromagnets by symmetry but uniquely allowed in altermagnets, highlighting a novel and intrinsic characteristic of altermagnetism. To systematically explore second-order spin current responses, we classify all symmetry-allowed responses based on spin point groups. Furthermore, we provide a comprehensive classification of altermagnetic Weyl semimetals by identifying spin space groups that host symmetry-enforced Weyl points. Utilizing this classification, we construct a symmetry-guided tight-binding model and confirm our predictions. Finally, we identify Weyl crossings in a material candidate via first-principle calculations. Our work unveils a distinctive optical response of altermagnets, paving the way for a new frontier in altermagnetism.
\end{abstract}

\maketitle

\textit{Introduction}---Spintronics~\cite{Zutic.RevModPhys2004,Bader.AnnRevCondMat2010,Baltz.RevModPhys2018,jungwirth.NatNanotech2016} utilizes spin degree of freedom in addition to charge, to store, process, and transfer information. Until now, there are several approaches to generate spin currents such as spin galvanic effect~\cite{edelstein.1990,ganichev.Nature2002}, spin Hall effect~\cite{Dyakonov1971,Hirsch1999,Murakami2004,Sinova2004,Kato2004,Wunderlich2005}, spin Seebeck~\cite{uchida2008,Uchida2010,jaworski2010,Xiao2010} and spin Nernst effect~\cite{cheng2008,liu2010,tauber2012,wimmer2013,meyer2017,sheng2017}. Among these, the generation of pure spin currents--currents composed solely of spin without an accompanying charge current--is particularly intriguing. Recently, optical generation of spin currents in antiferromagnets (AFMs) using bulk photovoltaic effects (BPVEs)~\cite{young_prediction_2013,xiao_spin_2021,fei_pt-symmetry-enabled_2021,xu_pure_2021,xiao2023} have attracted interest. This method does not rely on the spin-orbit coupling unlike other optical generation of spin currents in non-magnetic materials~\cite{bhat_optically_2000,bhat_pure_2005,zhao_injection_2005} and is promising due to its ultra-fast response. However, in AFMs, the generation of pure spin currents requires mirror symmetry to decouple spin and charge currents. This constraint significantly limits the pool of viable materials for practical applications.

\begin{figure}[h]
    \includegraphics[width=\columnwidth]{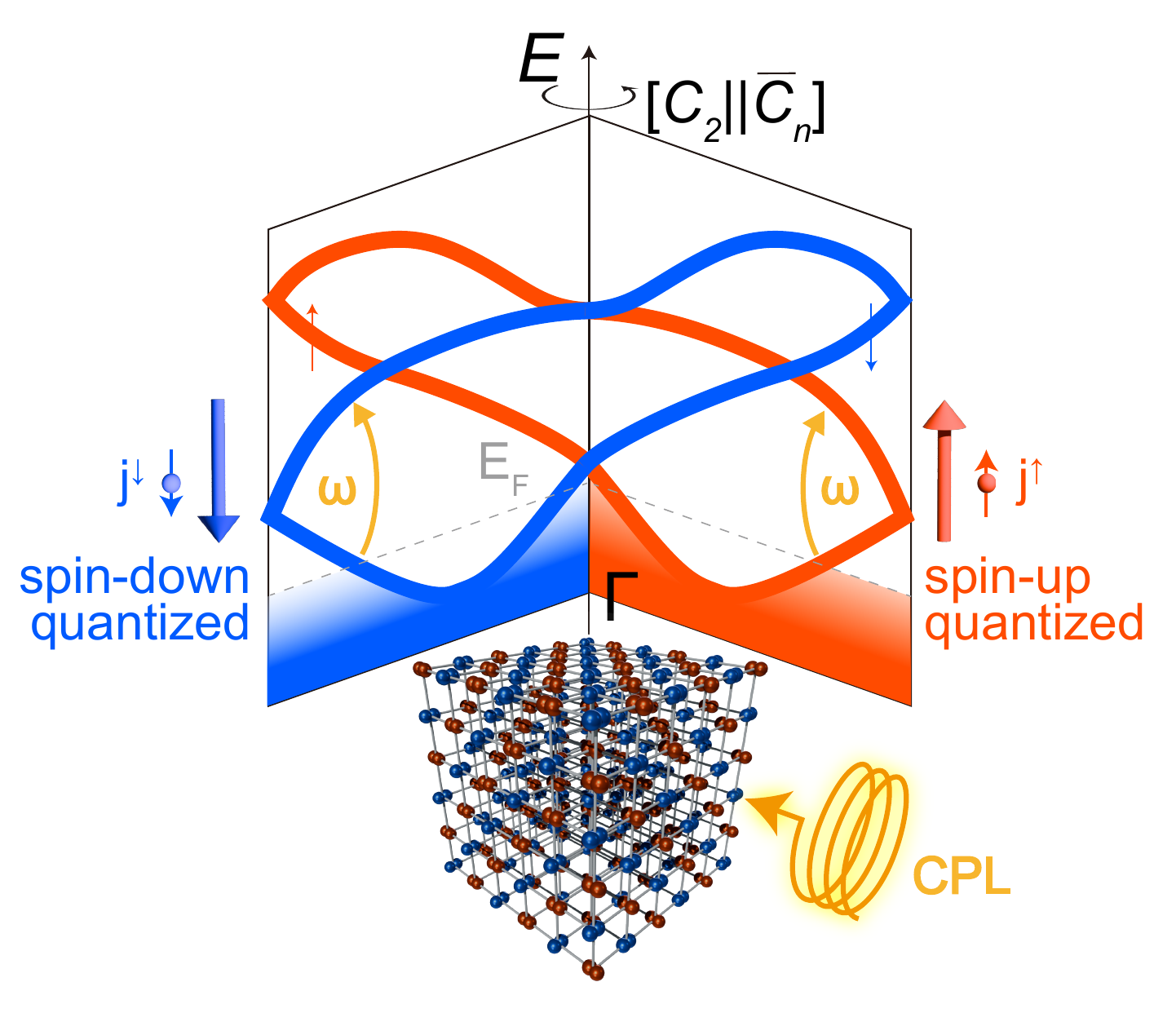}
    \caption{Schematic illustration of quantization of spin CPGE. Spin-polarized Weyl points in altermagnetic Weyl semimetals create quantized spin-polarized photocurrent. Oppositely polarized currents flow in the opposite direction by crystal symmetry and produce quantized pure spin currents.}
    \label{fig:schematics}
\end{figure}

The BPVE itself is also an intriguing research field. The BPVE refers to the generation of a direct current under illumination in non-centrosymmetric bulk materials~\cite{kraut.vonbaltz1979a,belinicher.sturman1980,vonbaltz.kraut1981a,aversa.sipe1995a,sipe.shkrebtii2000,fridkin2001}. Unlike conventional interface-based photovoltaics, the photovoltage produced by this effect is not constrained by the bandgap of materials, making the BPVE a promising candidate for realizing efficient next-generation solar cells. Moreover, the BPVE is known to be closely linked to the quantum geometrical properties of a material~\cite{morimoto.nagaosa2016,morimoto.nagaosa2016b,ma.etal2021,ahn.etal2022,hsu.etal2023b,morimoto.etal2023b,Jankowski2024,Alexandradinata2024,Jankowski2024b,avdoshkin2025multistategeometryshiftcurrent,mitscherling2025gaugeinvariantprojectorcalculusquantum}. One manifestation is the quantization of the circular photogalvanic effect (CPGE) in Weyl semimetals~\cite{dejuan.etal2017a}. As the CPGE is a second-order optical effect, generated photocurrent can be written as
\be
    j_a=\beta_{ab}\qty[\vb{E}(\omega)\times\vb{E}^*(\omega)]_b.\label{eq:CPGE}
\ee
It was shown that the trace of the conductivity $\beta_{ab}$ is proportional to the surface integral of the Berry curvature over a closed surface in $\vb{k}$-space where the frequency of the light $\omega$ matches the energy difference between initial and final states during optical excitation. If this surface encloses Weyl points, $\mathrm{tr}[\beta]$ is quantized in unit of $i\beta_0=i\pi e^3/h^2$. This follows from the fact that the surface integral of the Berry curvature around a Weyl point is equal to its monopole charge, which is a topological quantity and takes integer values. However, if two oppositely charged Weyl points lie at the same energy, their contributions to the surface integral cancel, making $\mathrm{tr}[\beta]=0$. Since mirror planes enforce equal energies for oppositely charged Weyl points, the absence of mirror planes is a critical requirement for the quantization of CPGE.

In this letter, we systematically and comprehensively identify that a subclass of recently discovered altermagnets~\cite{smejkal_beyond_2022,smejkal_emerging_2022} that can generate pure spin currents based on spin group symmetry. Altermagnets show spin-split bands and compensating magnetization, combining characteristics of both ferromagnets and antiferromagnets. Due to its time-reversal symmetry breaking and d-, g-, or i-wave form of non-relativistic spin-splitting, altermagnets are expected to show unique phenomena such as anomalous Hall effect~\cite{smejkal_emerging_2022,smejkal_crystal_2020,feng_anomalous_2022,gonzalez_betancourt_spontaneous_2023,wang_emergent_2023,jin_anomalous_2024,reichlova_observation_2024}, spin-splitter effect~\cite{gonzalez-hernandez_efficient_2021,karube_observation_2022,bai_observation_2022,bai_efficient_2023,giil_quasiclassical_2024,guo_direct_2024}, and distinctive optical responses~\cite{adamantopoulos_spin_2024}. Their realization in real materials is also the subject of active investigation~\cite{noda_momentum-dependent_2016,okugawa_weakly_2018,ahn_antiferromagnetism_2019,naka_spin_2019,reimers_direct_2024,reichlova_observation_2024,hariki_x-ray_2024}. Very recently, BPVE in altermagnets has been explored in some studies~\cite{Dong_etal_2024,Ezawa_2024}. However, systematic investigations based on spin group symmetries remain lacking.

\begin{figure}[t]
    \begin{center}
        \includegraphics[width = \columnwidth]{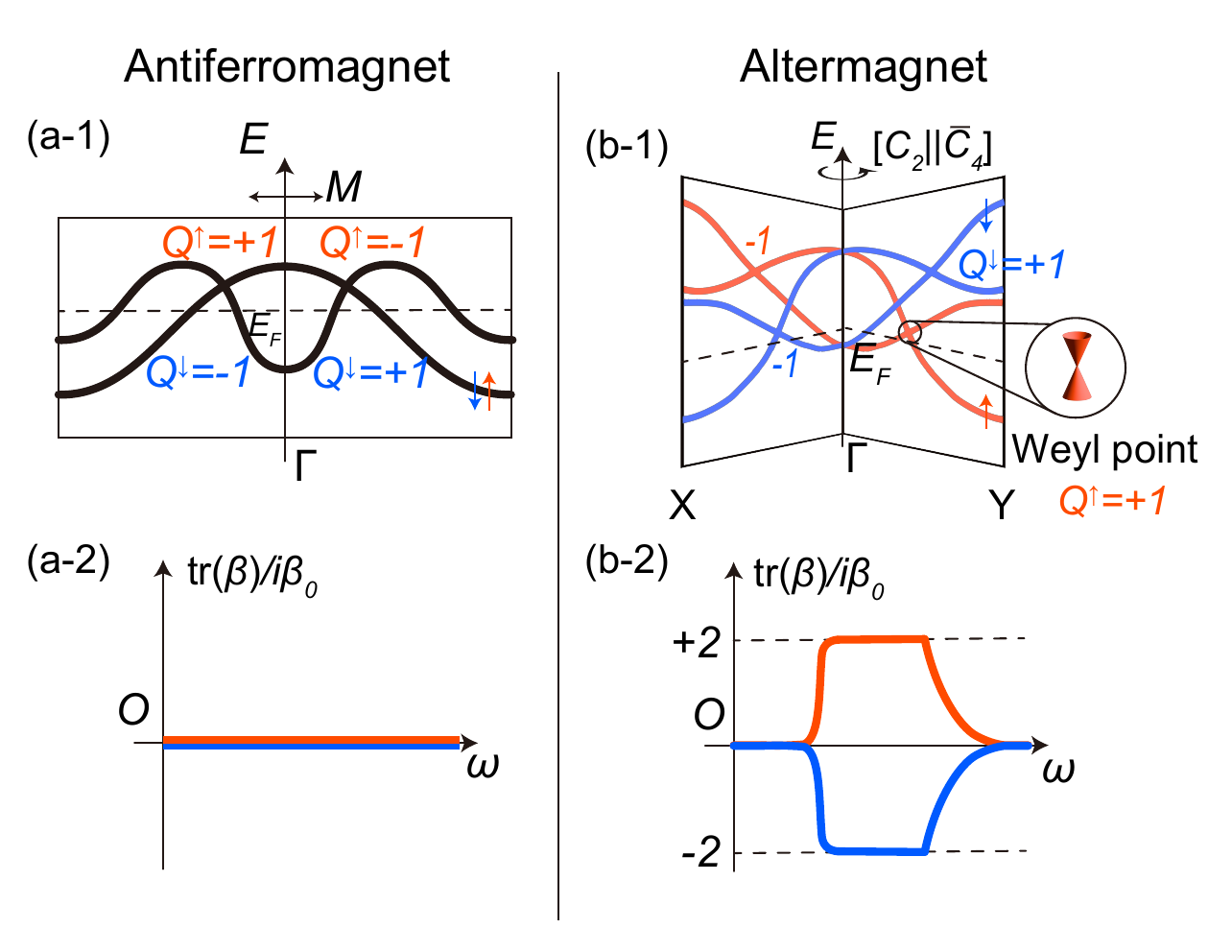}
        \caption{Comparison of spin CPGE in antiferromagnets and altermagnets. (a-1) Band structure of $\mathcal{PT}$-symmetric antiferromagnetic Weyl semimetals with a mirror plane. (a-2) The vanished trace of the spin conductivities $\mathrm{tr}[\beta^{\uparrow,\downarrow}]$. The red curve shows $\mathrm{tr}[\beta^{\uparrow}]$ and the blue for the spin down, respectively. (b-1) Spin-split energy bands of an altermagnet with $[C_2||\bar{C}_4]$ symmetry as an example. The Weyl points lie along $\Gamma\textrm{-} X$ and $\Gamma\textrm{-} Y$, and also along $\Gamma\textrm{-}\bar{X}$ and $\Gamma\textrm{-}\bar{Y}$ (not shown). (b-2) The quantized trace of the spin conductivity for altermagnetic case. The value is $+2$ reflecting the sum of monopole charges along $\Gamma\textrm{-} Y$ and \(\Gamma\textrm{-}\bar{Y}\) lines for the spin-up sector.}
        \label{fig:Schematic_AMAFM}
    \end{center}
\end{figure}

As a unique phenomenon of altermagnets, we also propose the spin-current analog of the quantization of CPGE (Fig.~\ref{fig:schematics}). As noted above, in order to generate a pure spin current in AFM, we need a mirror plane. However, by the combination of $\mathcal{PT}$- and mirror symmetries, oppositely charged Weyl points appear at the same energy for each spin (Fig.~\ref{fig:Schematic_AMAFM} (a-1)). Therefore, $\mathrm{tr}[\beta^{\mathrm{spin}}]\coloneqq\mathrm{tr}[\beta^{\uparrow}]-\mathrm{tr}[\beta^{\downarrow}]$ is zero, and spin conductivity vanishes in AFMs (Fig.~\ref{fig:Schematic_AMAFM} (a-2)). We can avoid this problem in altermagnets. For example, a combination of spin rotation and four-fold rotoinversion $[C_2||\bar{C}_4]$ does not force oppositely charged Weyl points to be at the same energy (Fig.~\ref{fig:Schematic_AMAFM} (b-1)). Here, $[S||R]$ consists of an operation $S$ in the spin-only space and an operation $R$ in the real space. In this case, we expect $\mathrm{tr}[\beta^{\uparrow}]$ and $\mathrm{tr}[\beta^{\downarrow}]$ be quantized with opposite signs, resulting a quantized pure spin CPGE (Fig.~\ref{fig:Schematic_AMAFM} (b-2)). To validate our scenario, we first show that the pure spin BPVE is possible in altermagnets by symmetry argument and then see the quantization of the pure spin CPGE. We also propose a material for the realization of this quantization using first-principle calculations. To date, CrSb is the only altermagnetic Weyl metal confirmed experimentally~\cite{Li_2025_CrSB,Lu_2025_CrSb}. Altermagnetic topological semimetals are still under active investigation~\cite{sunDiracNodalLines2017,jovicDiracNodalLines2018,smejkalCrystalTimereversalSymmetry2020a,zhouCrystalThermalTransport2024b}, and the phenomena predicted by our theory provide additional motivation and guidance for future experimental and theoretical studies in this field.

\textit{Spin bulk photovoltaic effects in altermagnets}---We start by investigating what kind of second-order optical responses are allowed in general altermagnets. Throughout this letter, we focus exclusively on collinear altermagnets. In this case, these systems have symmetry with respect to the semi-direct product of SO(2), the continuous symmetry about rotations around an axis parallel to the direction of spins, and the binary group $\mathbb{Z}_2=\left\{E, [\bar{C}_2||\mathcal{T}]\right\}$, where $\bar{C}_2$ is the spin rotation around an axis perpendicular to spins followed by an inversion operation in the spin-only space. In addition to these symmetries, altermagnets are described by spin point groups (SPGs), composed of spin-rotation and real-space point group operations~\cite{litvinopechowski1974,litvin1977}.

We now consider the bulk photovoltaic effect (BPVE). The BPVE is a phenomenon where a direct current is induced by light in non-centrosymmetric bulk crystals, and it is a second-order optical response expressed as
\be
    j^s_a&=\sigma^s_{abc}(\omega)E_b(\omega)E_c(-\omega),\label{eq:BPVE_formula}
\ee
where $a,b,c$ denote directions in real space, and $s$ denotes the spin polarization. As we discuss non-relativistic collinear magnets with $\mathrm{SO}(2)$-symmetry, spin is a good quantum number $s=\uparrow,\downarrow$ and we can define the spin current as
\be
    j_a^{\mathrm{spin}}\coloneqq j^{\uparrow}_a-j^{\downarrow}_a.
\ee
The BPVE is known to consist of two types of currents: the shift current, which is a polarization current, and the injection current, which is a dissipative current. When we decompose the conductivity tensor into real and imaginary parts as $\sigma^s_{abc}=\sigma^s_{L,abc}+i\sigma^s_{C,abc}$, with real $\sigma^s_{L,abc}$ and $\sigma^s_{C,abc}$ we can see $\sigma^s_{L,abc}=\sigma^s_{L,acb}$ and $\sigma^s_{C,abc}=-\sigma^s_{C,acb}$, from Eq.~\eqref{eq:BPVE_formula}. The real part $\sigma^s_{L,abc}$ contributes to currents regardless of the polarization of the light and the imaginary part $\sigma^s_{C,abc}$ contributes only to the circularly polarized light. Following Ref.~\cite{Ahn_divergence_2020}, we call $\sigma^s_{L,abc}$ linear conductivity and $\sigma^s_{C,abc}$ circular conductivity. Depending on the combination of currents (injection or shift) and the polarization of light (linear or circular), signs of the conductivity tensor under $\mathcal{T}$, $\mathcal{PT}$, and $[\bar{C}_2||\mathcal{T}]$ operations vary and the results are summarized in Tab.~\ref{tab:response}. For collinear magnets, the system possesses the aforementioned $\mathrm{SO}(2)\rtimes\mathbb{Z}_2$ symmetry and the response is restricted to circular injection and linear shift currents (For derivation, see Sec.~I of the supplemental material~\cite{supplementary}).

\begin{table}[t]
    \centering
    \caption{Signs of shift and injection electric (ele) and spin currents for linearly polarized light (LPL) and circularly polarized light (CPL) under time-reversal $(\mathcal{T})$, combination of time-reversal and spacial inversion $(\mathcal{PT})$, and $[\bar{C}_2||\mathcal{T}]$ operations. If the system is symmetric under each operation, only currents with positive sign are allowed.}
    \label{tab:response}
    \begin{tabular}{|cc|c|c|c|c|}
        \hline
        && L-inj & C-inj & L-shift & C-shift \\ \hline
        \multirow{2}{*}{$\mathcal{T}$}&ele & $-$ & + & + & $-$ \\ \cline{2-6}
        & spin & + & $-$ & $-$ & + \\ \hline
        \multirow{2}{*}{$\mathcal{PT}$}&ele & + & $-$ & $-$ & + \\ \cline{2-6}
        & spin & $-$ & + & + & $-$ \\ \hline
        \multirow{2}{*}{$\left[\bar{C}_2||\mathcal{T} \right]$} & ele & $-$ & + & + & $-$ \\ \cline{2-6}
        & spin & $-$ & + & + & $-$ \\ \hline
    \end{tabular}
\end{table}

\begin{table*}[t]
    \centering
    \caption{Classification of 27 non-centrosymmetric and altermagnetic spin point groups based on the possibility of spin and electric CPGEs. The SPGs in the right column show electric CPGE while SPGs in the bottom row show spin CPGE. The left bottom group includes 10 SPGs which allow pure spin CPGE.}
    \label{tab:spin_electric_currents}
    \renewcommand{\arraystretch}{1.3}
    \begin{tabular}{|c|c|c|}
        \hline
        & $\mathrm{Tr}[\beta^{\mathrm{ele}}]=0$ & $\mathrm{Tr}[\beta^{\mathrm{ele}}]\neq0$ \\
        \hline
        $\mathrm{Tr}[\beta^{\mathrm{spin}}]=0$ &
        \begin{minipage}{5.5cm}
            \centering
            $\begin{array}{c}
            ^1m^2m^22,^24^1m^2m,^2\bar{4}^22^1m,\\
            ^26^1m^2m,^1\bar{6}^2m^22,^2\bar{6}^1m^22
            \end{array}$
        \end{minipage}
        &\begin{minipage}{5.5cm}
            \centering
            $\begin{array}{c}
            ^22,^22^22^12,^24,^14^22^22,^24^12^22,^1\bar{4}^22^2m,\\
            ^13^22,^26,^16^22^22,^26^12^22,^24^13^22
            \end{array}$
        \end{minipage}
        \\
        \hline
        $\mathrm{Tr}[\beta^{\mathrm{spin}}]\neq0$ &
        \begin{minipage}{5.5cm}
            \centering
            $\begin{array}{c}
            ^2m,^2m^2m^12,^2\bar{4},^14^2m^2m,^2\bar{4}^12^2m,\\
            ^13^2m,^2\bar{6},^16^2m^2m,^2\bar{6}^2m^12,^2\bar{4}^13^2m
            \end{array}$
        \end{minipage}&No SPGs\\
        \hline
    \end{tabular}
\end{table*}

In order to have a non-zero BPVE, we need to break the inversion symmetry. There are 27 altermagnetic SPGs which satisfy this condition~\cite{litvin1977,smejkal_beyond_2022}. For all of them, we examine the form of the spin conductivity tensor. Under the application of the operation $[S||R]$ to the system, Eq.~\eqref{eq:BPVE_formula} becomes
\be
    j^{s'}_{a'}=([S||R]^{-1})^{s's}_{a'a}\sigma^s_{abc}[E||R]_{bb'}[E||R]_{cc'}E_{b'}E_{c'}.
\ee
When the system is invariant under the operation $[S||R]$, it means that
\be
    \sigma^{s'}_{a'b'c'}=([S||R]^{-1})^{s's}_{a'a}\sigma^s_{abc}[E||R]_{bb'}[E||R]_{cc'}.\label{eq:SigmaCondition}
\ee
We can get the spin-dependent conductivity tensor for each spin group by listing all conditions by using all elements of the SPG and solving the equations given by Eq.~\eqref{eq:SigmaCondition}. In the supplemental material~\cite{supplementary}, we show all components of spin-dependent conductivity tensors. The conductivity tensor for the spin current is $\sigma^{\mathrm{spin}}_{abc}\coloneqq\sigma_{abc}^{\uparrow}-\sigma_{abc}^{\downarrow}$. Similarly, the electric current is given by $j_a^{\mathrm{ele}}\coloneqq j_a^{\uparrow}+j_a^{\downarrow}$ and the conductivity tensor for the electric current is $\sigma^{\mathrm{ele}}_{abc}\coloneqq\sigma_{abc}^{\uparrow}+\sigma_{abc}^{\downarrow}$. We also list these conductivities in~\cite{supplementary}. If the component of $\sigma^{\mathrm{spin}}_{abc}$ is non-zero but $\sigma^{\mathrm{ele}}_{abc}$ is zero, then we can generate the pure spin current in the direction $a$ by the corresponding electric fields. We can see that, unlike in antiferromagnets, the existence of mirror planes is not a prerequisite for the generation of pure spin currents in altermagnets. These results apply to second-order conductivity tensors in all non-centrosymmetric altermagnets, even in the absence of Weyl points.

\textit{Quantization of pure spin circular photogalvanic effect}---Here we see that the quantization of pure spin CPGE is possible in altermagnets. For the CPGE, the spin conductivity $\beta^s_{ab}$ and the circular conductivity $\sigma^s_{C,abc}$ are related as
\be
    \beta^s_{ab} &=\frac{1}{2}\sum_{ij}\varepsilon_{bij}\sigma^s_{C,aij},
\ee
for each spin. Then, $\mathrm{tr}[\beta^{s}]=\sigma_{xyz}^{s}+\sigma_{yzx}^{s}+\sigma_{zxy}^{s}$ and we need SPGs which have $\mathrm{tr}[\beta^{\mathrm{spin}}]\neq0$ and $\mathrm{tr}[\beta^{\mathrm{ele}}]=0$ for the pure spin CPGE. In Tab.~\ref{tab:spin_electric_currents}, we classify 27 SPGs into four groups depending on the possibility of generating electric and spin CPGE. The following 10 SPGs can generate the pure spin CPGE.
\be
    \begin{split}
        \prescript{2}{}{m},\,\prescript{2}{}{m}^2m^12,\,\prescript{2}{}{\bar{4}},\, \prescript{1}{}{4}^2m^2m,\,\prescript{2}{}{\bar{4}}^12^2m,\\
        \prescript{1}{}{3}^2m,\, \prescript{2}{}{\bar{6}},\,\prescript{1}{}{6}^2m^2m,\, \prescript{2}{}{\bar{6}}^2m^12,\,\prescript{2}{}{\bar{4}}^13^2m\label{eq:list_SPGs}
    \end{split}
\ee
This is one of our main results. This list contains all types of spin-momentum locking in altermagnets, which are combinations of planar or bulk and d-, g- or i-wave types.

\textit{Altermagnetic Weyl semimetals}---Here we identify spin space groups (SSGs) with symmetry protected Weyl points. In general, SSG $\vb{F}_{\mathrm{S}}$ is constructed as~\cite{litvinopechowski1974}
\be
    \vb{F}_{\mathrm{S}} = \qty[E||\vb{f}]+\qty[C_2||\vb{F}-\vb{f}],\label{eq:SSG_coset}
\ee
where we call space groups $\vb{F}$ and $\vb{f}$ as parent and single-spin space groups, respectively, as elements of $\qty[E||\vb{f}]$ does not alter the spin. For the quantization of the pure spin CPGE, we need Weyl points composed of bands with the same spin polarization. Thus, we need to consider the symmetry of $\vb{f}$ which enforces Weyl points.

In addition to $\vb{f}$, the $\qty[\bar{C}_2||\mathcal{T}]$-symmetry of collinear magnets plays an important role here. As the $\bar{C}_2$ operation in the spin-only space keep the spin but the $\qty[E||\mathcal{T}]$ operation acts in the real space as a time-reversal operation, the $\qty[\bar{C}_2||\mathcal{T}]$-symmetry can be thought as an effective time-reversal symmetry which does not flip the spins.

Degeneracies at high-symmetry points in the Brillouin zone for a given space group are enforced by symmetry and they can be seen from the irreducible representations of the electronic bands. For time-reversal symmetric systems, Ref.~\cite{Yu_encyclopedia_2022} lists all symmetry-enforced band degeneracies in all space groups. Among them, we choose space groups $\vb{f}$ with Weyl points. Then, to make an altermagnetic SSG $\vb{F}_{\mathrm{S}}$, the super group of $\vb{f}$ should have a two-coset decomposition satisfying Eq.~\eqref{eq:SSG_coset}. Finally, we pick up the SSG $\vb{F}_{\mathrm{S}}$ such that the corresponding SPG is in our list of SPGs that exhibit quantization of the pure spin CPGE (Eq.~\eqref{eq:list_SPGs}). Based on these criteria, we find 34 SSGs which host symmetry-enforced Weyl points and compatible with Eq.~\eqref{eq:list_SPGs}. One example of such SSG is $R\prescript{1}{}{3}^2c$ and we construct a tight-binding model of this SSG below to see the quantization. In the Supplemental material~\cite{supplementary}, we provide a full list of the SSGs. These SSGs are optimal candidates in the sense that they necessarily host Weyl points enforced by symmetry; however, it is important to note that they are not prerequisites for quantization. Other SSGs corresponding to the SPG in Eq.~\eqref{eq:list_SPGs} may also exhibit quantization if they host accidental Weyl points.

\begin{figure}[t]
    \begin{center}
        \includegraphics[width = \columnwidth]{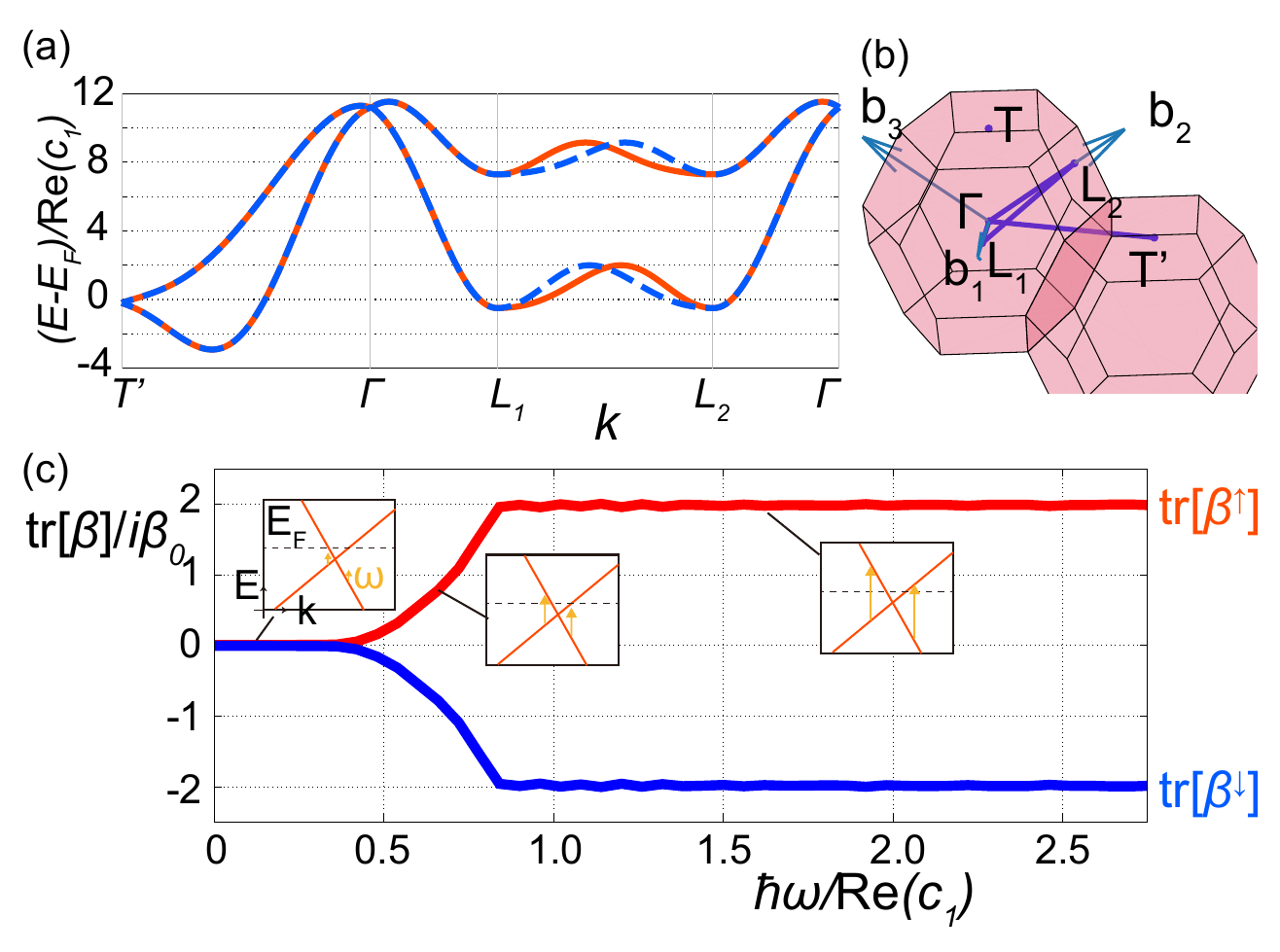}
        \caption{Calculation of the conductivity tensor for a tight-binding model of SSG $R\prescript{1}{}{3}^2c$. (a) Band structure of the model. Red solid lines indicate spin-up bands and blue dashed lines indicate spin-down bands. (b) The high-symmetry points and the path are shown in the top view of the reciprocal space. The points $\Gamma=\vb{0},\,L_1=\vb{b}_1/2,\,L_2=\vb{b}_2/2,$ and $T'=T-\vb{b}_3=(\vb{b}_1+\vb{b}_2-\vb{b}_3)/2$ lie on the glide plane, where the bands are spin-degenerate. Here, reciprocal vectors are $\vb{b}_1=2\pi\qty(1/a,1/(\sqrt{3}a),1/c),\vb{b}_2=2\pi\qty(-1/a,1/(\sqrt{3}a),1/c)$, and $\vb{b}_3=2\pi\qty(0,-2/(\sqrt{3}a),1/c)$. Hopping parameters are $(c_1,c_2,c_3,c_4)=(1+1.5i,1.9+1.6i,1.3+0.53i,0.83+0.67i)$, the lattice constants are $a=1,\,c=6$, and the Fermi energy is $E_F=-2.5$, ensuring that the Weyl point at $T'$ is close to the Fermi energy. (c) Plots of the spin-polarized traces of the conductivity tensors $\mathrm{tr}[\beta^{\uparrow,\downarrow}]$, in units of $i\beta_0=ie^3\pi/h^2$ as functions of the light frequency $\omega$. The insets show schematic photo-excitation around a Weyl point.}
        \label{fig:ResultsTBM}
    \end{center}
\end{figure}

\textit{Model calculation}---Here we construct a tight-binding model of the SSG $R\prescript{1}{}{3}^2c$ as an example and numerically calculate the conductivity. As $R\prescript{1}{}{3}^2c=\qty[E||R3]+\qty[C_2||R3c-R3]$, Hamiltonians for spin-up bands $H^{\uparrow}$ and spin-down bands $H^{\downarrow}$ both obey symmetries of the space group $R3$ and the effective time-reversal symmetry $\qty[\bar{C}_2||\mathcal{T}]$. Two Hamiltonians $H^{\uparrow}$ and $H^{\downarrow}$ are related by symmetry operations in $R3c-R3$. By placing $p_x$ and $p_y$ orbitals at Wyckoff position 3a, we get two-dimensional irreducible representations at the $\Gamma$ and $T$ points~\cite{Bilbao1,Bilbao2}. We consider the nearest-neighbor hoppings between these orbitals and the tight-binding Hamiltonian in the basis $(p_x+ip_y,p_x-ip_y)$ can be written as (see \cite{supplementary} for details)
\be
    H^{\uparrow}(\vb{k})&=\mqty(h_{++}(\vb{k})&h_{-+}(\vb{k})\\h_{+-}(\vb{k})&h_{--}(\vb{k})),\\
    H^{\downarrow}(\vb{k})&=\mqty(h_{--}(\vb{k}')& h_{+-}(\vb{k}')\\ h_{-+}(\vb{k}')& h_{++}(\vb{k}')), \\
    \vb{k}'&\coloneqq (-k_x,k_y,k_z),\\
    H(\vb{k})&=\mqty(H^{\uparrow}(\vb{k})&0\\0&H^{\downarrow}(\vb{k})),
\ee
where
\be
    h_{++}(\vb{k})&\coloneqq \sum_{j=1}^3\Re\qty(c_1e^{-i\vb{k}\cdot\vb{d}_j}+c_2e^{-i\vb{k}\cdot\vb{a}_j})\\
    h_{--}(\vb{k})&\coloneqq \sum_{j=1}^3\Re\qty(c_1^*e^{-i\vb{k}\cdot\vb{d}_j}+c_2^*e^{-i\vb{k}\cdot\vb{a}_j})\\
    h_{+-}(\vb{k})&\coloneqq \sum_{j=1}^3e^{i\frac{2}{3}\pi(j-1)}\qty(c_3\cos(\vb{k}\cdot\vb{d}_j)+c_4\cos(\vb{k}\cdot\vb{a}_j))
\ee
with $\vb{a}_1=(a/2,\sqrt{3}a/6,c/3)$, $\vb{a}_2=(-a/2,\sqrt{3}a/6,c/3)$, and $\vb{a}_3=(0,-\sqrt{3}a/3,c/3)$ are the primitive lattice vectors $(a,c\in\mathbb{R})$, $\vb{d}_1=\vb{a}_1-\vb{a}_2$, $\vb{d}_2=\vb{a}_2-\vb{a}_3$, $\vb{d}_3=\vb{a}_3-\vb{a}_1$, and complex hopping amplitudes $c_i \in \mathbb{C}$, $i=1,2,3,4$. The band structure of this system is shown in Fig.~\ref{fig:ResultsTBM} (a) along a path indicated in Fig.~\ref{fig:ResultsTBM} (b), where $\vb{b}_i$ are reciprocal lattice vectors corresponding to $\vb{a}_i\,(i=1,2,3)$. Spin-up (red solid curves) and spin-down (blue dashed curves) bands are degenerate on glide planes such as $k_x=0$. There are two Weyl points in the Brillouin zone, one at the $\Gamma$ point and the other at the $T$ point \(\left(\vb{k}=\frac{1}{2}\left(\vb{b}_1+\vb{b}_2+\vb{b}_3\right)\right)\) for each spin sector. The Weyl points at $T$ with energy $(E-E_F)/\Re(c_1)=-0.2$ carry monopole charges $Q^{\uparrow}=+2$ and $Q^{\downarrow}=-2$ and those at \(\Gamma\) are with \(Q^{\uparrow}=-2\) and \(Q^{\downarrow}=+2\).

The CPGE is the injection current response to the circularly polarized light. The injection conductivity tensor can be written as~\cite{Ahn_divergence_2020}
\be
    \sigma^{\mathrm{inj}}_{abc}&=-\tau\frac{2\pi e^3}{\hbar^2}\int_{\vb{k}}\sum_{n,m}f_{nm}^{\mathrm{FD}}\Delta_{mn}^ar^c_{nm}r^b_{mn}\delta\qty(\omega_{mn}-\omega),\label{eq:FormulaInj}
\ee
where $-e\ (e>0)$ is the charge of an electron, $\int_{\vb{k}}=\int\mathrm{d}^dk/(2\pi)^d$ for a $d$-dimensional system, $f^{\mathrm{FD}}_{nm}$ is the difference of the Fermi-Dirac distribution functions of the bands $n$ and $m$, $f^{\mathrm{FD}}_{nm}=f^{\mathrm{FD}}_{n}-f^{\mathrm{FD}}_{m}$, $H\ket{n}=\hbar\omega_n\ket{n}$, $v^a_{nm}=\hbar^{-1}\bra{n}\partial_{k_a}\ket{m}$, $\hbar\omega_{mn}=\hbar\omega_m-\hbar\omega_n$ is the energy difference between two bands $m$ and $n$, and $r^c_{nm}=\bra{m}i\partial_{k_c}\ket{n}$ is a component of the Berry connection. Here, $\Delta_{mn}^a=v^a_{mm}-v^a_{nn}$ is the velocity difference between two bands $n$ and $m$. We can calculate $\mathrm{tr}[\beta^s]$ using this formula.

For this system, $\mathrm{tr}[\beta^{\uparrow,\downarrow}]$ are plotted in Fig.~\ref{fig:ResultsTBM} (c). As expected, conductivities for spin up and down have opposite signs and show quantization $\mathrm{tr}[\beta^{\uparrow,\downarrow}]/i\beta_0=\pm2$. From $\hbar\omega/\Re(c_1)=0.4$ to $0.8$, $\mathrm{tr}[\beta^s]$ changes gradually from $0$ to $\pm2$. This gradual change is due to the anisotropic shape of the Weyl point. At $\hbar\omega/\Re(c_1)=0.4$, the photo-excitation at some $\vb{k}$ around the Weyl point at the $T$ point occur and at $0.8$, the integration surface become closed resulting the perfect quantization.

\begin{figure}[t]
    \begin{center}
        \includegraphics[width = \columnwidth]{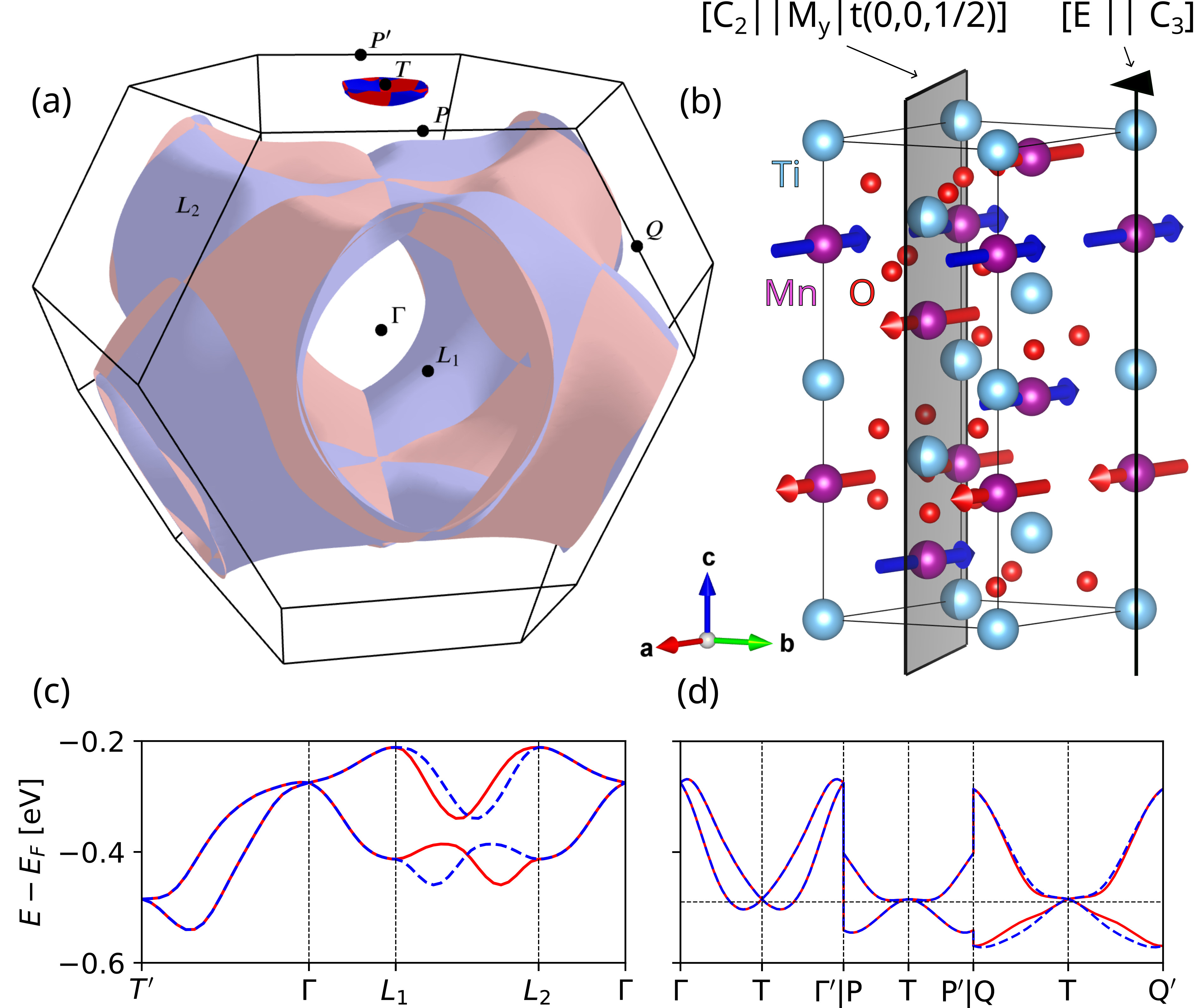}
        \caption{DFT calculation of $\mathrm{MnTiO_3}$ without spin-orbit coupling. Red and blue lines (surfaces) show spin-up and spin-down bands, respectively. (a) Constant-energy isosurface at $E = E_F - 0.49~\mathrm{eV}$. Highlighted are the topmost valence bands, converging to the quadratic Weyl point $T$. Time-reversal symmetry is broken and spin-splitting with even-parity wave (g-wave) anisotropy is visible. (b) Conventional (hexagonal) unit cell of $\mathrm{MnTiO_3}$. Spin space group generators are shown: mirror $\mathrm{[C_2||M_y|t(0, 0, \frac{1}{2})]}$ connects sites with opposite spins, 3-fold rotation $\mathrm{[E||C_3]}$ connects same-spin sites. (c,d) Band structures of 4 topmost valence bands. Altermagnetic spin splitting of roughly 50 meV is visible. (c) Path equal to the model band structure. (d) Path along 3 lines around $T = (1/2, 1/2, 1/2)$. Horizontal line shows the energy level of the isosurface in (a). }
        \label{fig:ResultsDFT}
    \end{center}
\end{figure}

\textit{Material candidate}---Here we propose MnTiO$_3$ as a candidate material. The structure, shown in Fig.~\ref{fig:ResultsDFT} (b), belongs to the space group $R3c$. The magnetic moments on the Mn atoms alternate between $+x$ and $-x$ direction -- the magnetic structure is altermagnetic and belongs to the $R\prescript{1}{}{3}^2c$ spin space group, implying g-wave spin-splitting and 4 nodal surfaces in the band structure. The magnetic dipolar structure was reported by neutron diffraction, with lattice parameters $a = 5.187$ \AA, $c = 13.680$ \AA~\cite{MnTiO3_structure}. We performed first-principle calculation using density functional theory (DFT) via the Vienna Ab Initio Simulation Package (\textsc{VASP}) \cite{VASP1, VASP2, PAW} in the collinear magnetism setting (without spin-orbit interaction) with the Perdew-Burke-Ernzerhof (PBE) exchange-correlation functional \cite{PBE}. Primitive unit cell containing 2 Mn atoms was used in the calculation. Further information about the calculation is available in Sec.~5 of the supplemental material~\cite{supplementary}.

Figure~\ref{fig:ResultsDFT} (c) shows the band structure along the same path as Fig.~\ref{fig:ResultsTBM} (c): the bands are spin-degenerate on the reciprocal high-symmetry paths $\Gamma$-($L_1$, $L_2$, $T'$) and spin-split on the $L_1$-$L_2$ path, and quadratic dispersion is visible around $\Gamma$ and $T'$, same as in the model band structure.

The single-spin space group in the material $R3$ enforces quadratic Weyl points (QWP) with $|Q| = 2$ at $\Gamma$ and $T'$ between same-spin bands~\cite{Yu_encyclopedia_2022}, as shown in previous section.
The constant-energy isosurface in Fig.~\ref{fig:ResultsDFT} (a) and the band structure Fig.~\ref{fig:ResultsDFT} (d) shows the dispersion around the symmetry-enforced QWP at $T$. It is quadratic along two directions and linear along the $\Gamma - T$ direction.
The $\mathrm{[C_2||M_y]}$ symmetry enforces opposite-spin QWP at identical reciprocal-space point and energy. The two Weyl points have anisotropic dispersion fulfilling the $\mathrm{[E||C_3]}$ symmetry and are related by the mirror symmetry.
This results in alternating spin-splitting between the opposite-spin bands and 4 nodal planes.

In experiments, tuning the Fermi energy is necessary, and this can be achieved, for example, by doping or gating. However, the disorder introduced by doping can modify the quantized value of the CPGE~\cite{Wu_2024_disorder}. Unlike the usual CPGE, for the quantization of the spin CPGE, the pure spin current must be converted into an electric current before measurement, and this conversion inherently complicates the direct observation of the quantized value. Nevertheless, as long as the disorder does not disrupt the altermagnetic order or the Weyl points, the quantized plateau should still be observable.

\textit{Conclusion}---In this study, we demonstrated that the quantization of the spin circular photogalvanic effect can be achieved in altermagnetic Weyl semimetals. This effect cannot occur in conventional antiferromagnets, because the existence of the mirror plane, that is essential to generate pure spin currents in antiferromagnets, combined with the spin-degenerate bands cancel this effect. Thus, we provide a phenomenon unique to altermagnets. Through systematic analysis, we identified 10 out of 27 non-centrosymmetric altermagnetic spin point groups that support the quantization of the pure spin CPGE. Furthermore, we investigated spin space groups and tabulated 34 spin space groups that host symmetry-enforced Weyl points at high-symmetry points protected by crystalline symmetries and allows the quantization. From this list, we also proposed MnTiO$_3$ as a candidate material for an experimental realization with first-principle calculations. While the MnTiO3 is insulating, doping can bring the Weyl crossing points to Fermi level. These results provide strong support for the feasibility of experimental verification. Physical phenomena due to chiral anomaly are expected for these Weyl points. One may naively expect a pure spin current under parallel magnetic and electric fields, but it is not exactly the case in general, because the altermagnetic order may be broken by the magnetic field. Thus a nearly pure spin current can be expected in the weak magnetic field.

\textit{Acknowledgements}---HY acknowledges support by Japan Society for the Promotion of Science (JSPS) KAKENHI Grant Number JP24KJ1109 and by MEXT Initiative to Establish Next-generation Novel Integrated Circuits Centers (X-NICS) Grant Number JPJ011438. L\v{S} and JP acknowledge funding from the ERC Starting Grant No. 101165122.
SM acknowledges support by Japan Society for the Promotion of Science (JSPS) KAKENHI Grant Numbers JP22K18687, JP22H00108, and JP24H02231.

\bibliography{AM_SBPVE.bib}

\end{document}